\documentclass[prb,twocolumn,superscriptaddress,floatfix]{revtex4}
\usepackage{graphicx} 
\usepackage{amsmath}
\usepackage{upgreek}
\newcommand{\dd}{{\textrm{d}}}
\usepackage[dvips]{epsfig}
\newcommand{\beq}{\begin{equation}}
\newcommand{\eeq}{\end{equation}} 
\newcommand{\beqa}{\begin{eqnarray}}
\newcommand{\eeqa}{\end{eqnarray}}
\newcommand{\ba}{\begin{array}}
\newcommand{\ea}{\end{array}}

\begin{document}

\title{Merging of superfluid Helium nanodroplets with vortices}
\author{Jos\'e Mar\'{\i}a Escart\'{\i}n}
\affiliation{Catalan Institute of Nanoscience and Nanotechnology (ICN2),
CSIC and BIST, Campus UAB,
Bellaterra,
08193 Barcelona, Spain.}

\author{Francesco Ancilotto}
\affiliation{Dipartimento di Fisica e Astronomia ``Galileo Galilei''
and CNISM, Universit\`a di Padova, via Marzolo 8, 35122 Padova, Italy}
\affiliation{ CNR-IOM, via Bonomea, 265 - 34136 Trieste, Italy }

\author{Manuel Barranco}

\affiliation{Departament FQA, Facultat de F\'{\i}sica,
Universitat de Barcelona. Diagonal 645,
08028 Barcelona, Spain}
\affiliation{Institute of Nanoscience and Nanotechnology (IN2UB),
Universitat de Barcelona, 08028 Barcelona, Spain.}

\author{Mart\'{\i} Pi}
\affiliation{Departament FQA, Facultat de F\'{\i}sica,
Universitat de Barcelona. Diagonal 645,
08028 Barcelona, Spain}
\affiliation{Institute of Nanoscience and Nanotechnology (IN2UB),
Universitat de Barcelona, 08028 Barcelona, Spain.}

\begin{abstract} 
Within Density Functional Theory, we have investigated 
the coalescence dynamics of two superfluid helium nanodroplets hosting vortex 
lines in different relative orientations,
which are drawn towards each other by the Van der Waals mutual attraction.
We have found a rich phenomenology depending on how the vortex lines are oriented. 
In particular, when a vortex and anti-vortex lines are present in the merging 
droplets, a dark soliton develops at the droplet contact region, which eventually decays into vortex rings. 
Reconnection events are observed between the vortex lines or rings, leading to 
the creation of more vortices. 
Our simulations show the interplay 
between vortex creation and reconnections, as well as the effect of the droplet surface
which pins the vortex ends and, by reflecting short-wavelength  excitations produced by the interactions between vortices,
strongly affects the droplet final state. Additional vorticity is nucleated
in the proximity of surface indentations produced in the course of the dynamics, which in turn 
interact with other vortices present in the  droplets. These effects, obviously absent in the case of bulk 
liquid helium, show that the droplet surface may  act as a multiplier of vortex reconnections. 
The analysis of the energy spectrum shows that vortex-antivortex ring annihilation, as well as vortex-antivortex reconnections, yields 
roton bursts of different intensity.
\end{abstract} 
\date{\today}

\maketitle

\section{Introduction}

Droplet merging is a well established subject of investigation in classical  fluids as it arises in different fundamental and practical 
contexts.\cite{Egg99,Duc03,Wu04,Bar10,Gu11}  It involves viscous droplets, mostly drawn 
together by surface tension or Van der Waals mutual attraction. 
Experiments and models aim at determining e.g. how the radius of the contact region between drops grows as
a function of time after contact, or the final outcome of the interaction between two droplets
(merging in one single droplet or separation into two or several), which depends on the size, viscosity, 
morphology and relative velocity of the colliding droplets, see e.g. Refs. \onlinecite{Qia97,Men01,Pan05,Pan09,Hua19} and
references therein.

At variance, little is known on the coalescence of drops made of superfluids, for which viscosity is negligible. 
Experiments on magnetically levitated helium droplets at 0.7 K temperature  have been carried out some time ago by Maris 
and coworkers.\cite{Vic00}
Using a static magnetic field, drops of less than 1 cm radius were confined and made to collide at velocities as small as a few cm/s.  
We have recently simulated this experiment using nanoscopic He droplets within the time-dependent density functional theory (TDDFT)
approach.\cite{Esc19}  The droplets were initially at rest and drawn together by the mutual Van der Waals (VdW) long-range attraction.
It was shown that the radius of the contact region grows as the root square of time after contact, similarly as 
for low viscosity classical drops.\cite{Wu04,Egg99,Duc03} This scaling has been also found for the
collision of self-bound quantum droplets made of ultradilute Bose mixtures of cold gases.\cite{Fer19}

\begin{figure}[!]
\centerline{\includegraphics[width=1.0\linewidth,clip]{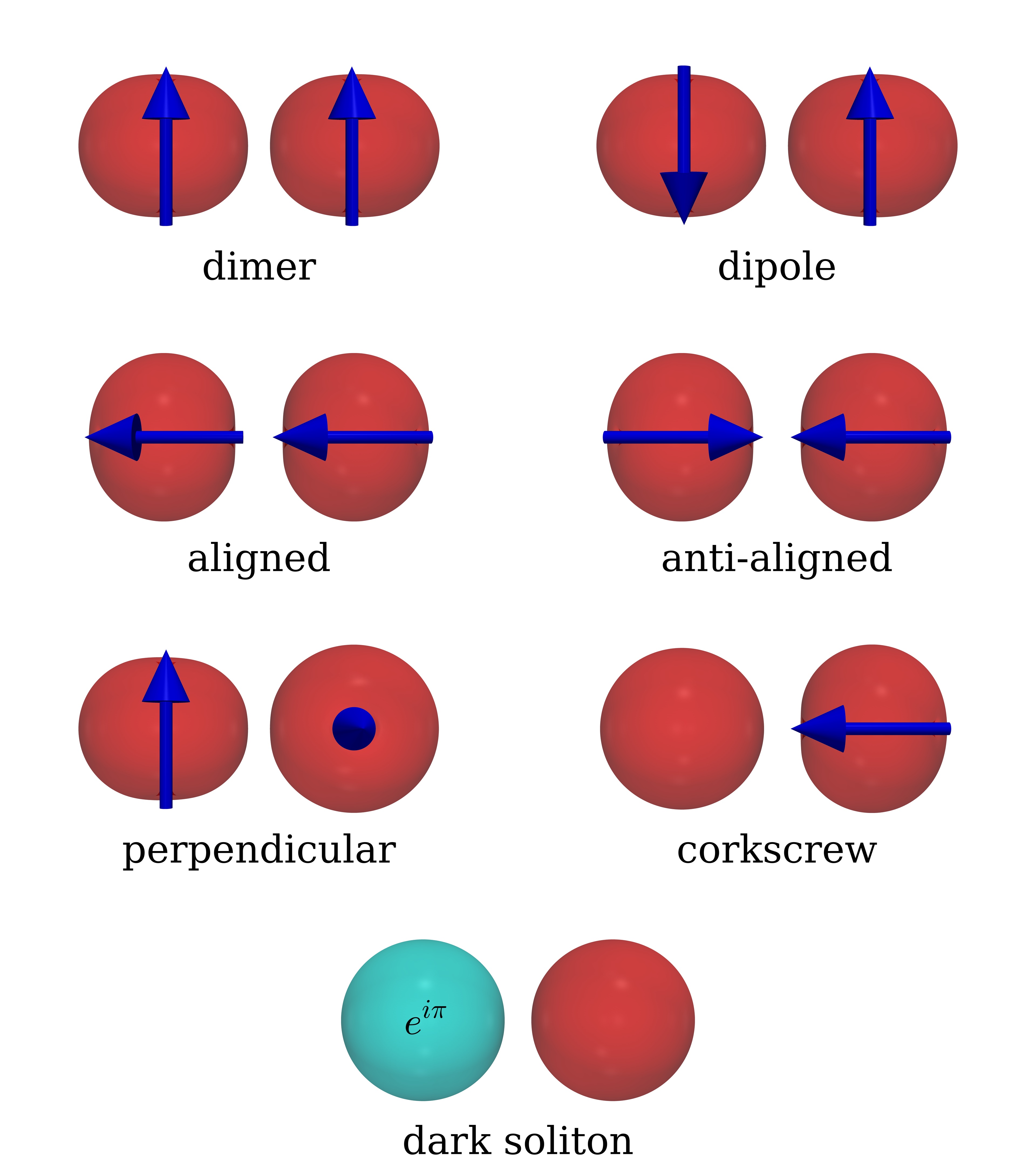}}
\caption{
Schematic view of the the merging droplets discussed in this work displaying the different orientation of the vortex lines they host.
}
\label{fig1}
\end{figure}

The merging of vortex-free helium droplets has unveiled the appearance of vortex-antivortex ring pairs nucleated at the droplet surface, that 
either wrap around the coalesced droplet or penetrated into it, eventually annihilating each other yielding an intense roton burst.\cite{Esc19}
These effects are most remarkable if one recalls that, initially, no vortices were  in the droplets and that droplets were not  in relative motion.
This calls for studying a more general situation where some vorticity is already present in the coalescent droplets, and constitutes the motivation
of the present work. Within TDDFT, we  address  here  the merging of two equal-size superfluid helium 
droplets at zero temperature, 
which may host vortex lines with different relative orientations, as schematically represented 
in Fig. \ref{fig1}.

We shall show that, depending on how  droplets are initially prepared, a rich variety of phenomena may occur, as
the merging process  changes the vorticity of the system  in a complex way.  
One important outcome of our simulations is a dynamical description of 
vortex interactions and reconnections, which are fundamental ingredients of current studies of quantum turbulence in liquid He and 
in cold-gas superfluids,   see e.g.  Refs. \onlinecite{Vin02,Tsu09,Bar14,Tsu17} for 
reviews on the subject.

Usually, numerical simulations of vortex reconnections in superfluids
start from bulk-like systems where vortex lines are  imprinted which, in order to initiate the 
vortex-vortex  interaction and ensuing reconnection, are subject to additional velocity fields which push them together.
Besides,  the actual microscopic structure of the vortex cores is  most often ignored, and vortex filament models are instead employed.
In the present simulations, interactions between vortices are  dynamically induced by the 
spontaneous merging  of the droplets; most
importantly, the droplet surface provides quite naturally an environment favoring both vortex interaction and proliferation.

This work is organized as follows. In Sec. II  we briefly describe the Density Functional Theory (DFT)
approach as applied to liquid $^4$He and  droplets. The results are discussed
in Sec. III, complemented with material presented in a Supplemental Material document;\cite{SM} a summary is presented in Sec. IV. 
Multimedia materials accompany this paper showing the real-time dynamics for the  merging processes addressed in this work.\cite{SM}
This material  constitutes an important part of this work, since often it is only by viewing how a complex process unfolds in time 
that one captures important physical details which would otherwise escape in a written account.

\section{Method}

The DFT approach, in its static and time-dependent formulations, has been probed to be 
 a realistic method  for studying a strongly-correlated  quantum fluid as liquid  helium;\cite{Anc17} it has been shown to 
sustain complex dynamical mechanisms, such as vortex nucleation and vortex-density waves interactions and, 
as shown in the following, vortex reconnections. The Gross-Pitaevskii (GP) approach\cite{Pit16} has been also shown 
to sustain these effects in dilute cold gases. In particular,
slow collisions of two confined Bose-Einstein condensates (BEC) in both ground and exited states,
have been addressed within this approach.\cite{Sun08}

Previous TDDFT simulations have  shown that impurities impinging nanoscopic superfluid He droplets nucleate
vortex rings and vortex loops that start and end on the droplet  surface.\cite{Mat14,Lea14} 
Other TDDFT simulations have addressed the capture of 
impurities by vortex-hosting droplets,\cite{Cop17} much as  it happens in the experiments where captured impurities 
form clusters along the vortex cores, helping detect the presence of vortices.\cite{Gom14,Oco20}

We provide here a basic account of the DFT approach  to help discuss the results presented in 
this paper, and refer the interested reader to Ref. \onlinecite{Anc17} for a detailed description of the method.

\subsection{Statics}

Within DFT, the total energy of a $^4$He$_N$ droplet   at zero temperature is written as a functional
of a complex effective wave function $\Psi( \mathbf{r})$ related to its atomic density by $\rho (\mathbf{r})= |\Psi( \mathbf{r})|^2$.
As a first step in the study of droplet merging, the  equilibrium configuration of an isolated helium droplet  is obtained by solving the 
static DFT equation for $\Psi( \mathbf{r})$
\begin{equation}
\left\{-\frac{\hbar^2}{2m} \nabla^2 + \frac{\delta {\cal E}_c}{\delta \rho}
\right\}\Psi({\mathbf r}) 
 \equiv
{\cal H}[\rho] \,\Psi({\mathbf r}) 
= \mu \Psi({\mathbf r})
\label{eq1}
\end{equation}
where $\mu$ is the $^4$He chemical potential and ${\cal E}_c[\rho]$ is the correlation energy. In this study, the number of helium atoms
in the  droplet  is $N=500$, and the results we present  have been obtained 
using the 4He-DFT BCN-TLS computing package.\cite{Pi17} Details on how Eq. (\ref{eq1})  
is solved can be found in  Refs. \onlinecite{Anc17,dft-guide} and references therein.

The correlation energy functional ${\cal E}_c[\rho]$ has been taken from  Ref.  \onlinecite{Anc05}. 
This functional is finite-range and includes non-local effects which are needed to 
describe quantitatively the response of the superfluid at  the scale of the vortex core radius,\cite{Gal14} 
providing an accurate description of the vortex core structure  and of the density modulations 
around it.\cite{Mat15a} These density modulations  have been  interpreted as a cloud of  bound, virtual rotons, embodied in the 
phase of the vortex wave function.\cite{Ame18} It has been furthermore suggested that these virtual 
rotons may be converted into real ones following vortex-antivortex annihilations, thus making vortex tangles 
a potential source of non-thermal rotons.\cite{Ame18} 

A vortex line along a diameter of the $^4$He$_{500}$ droplet can be easily generated by solving Eq. (\ref{eq1})
with a proper form for the starting wave function. For instance, the following choice   
\begin{equation}
\Psi_{start}(\mathbf{r})=\rho_0^{1/2}(\mathbf{r})\,
\left[ {x+\imath y \over \sqrt{x^2+y^2}}  \right]
\label{eq2}
\end{equation}
where $\rho_0(\mathbf{r})$ is the vortex-free  droplet density, yields upon convergence a vortex line along the $z$ axis with
positive (counter clock-wise) $n=1$ circulation number. A vortex line with $n=-1$ circulation number (antivortex) is instead
generated by taking in the numerator of Eq. (\ref{eq2}) the combination $x-\imath y$; 
vortices along the $x$ and $y$ axes can be generated by expressions similar to Eq. (\ref{eq2}).
These vortex states are excited droplet states, eigenstates of the angular momentum about the chosen diameter,
with eigenvalue $N \hbar$, i.e., $500 \hbar$ in the present case.  The total energy of the $^4$He$_{500}$ droplet is -2474.6 K, 
and that of the  $^4$He$_{500}$ droplet hosting a straight vortex line along a diameter is -2381.8 K, corresponding to an excitation energy of 92.7 K.

\subsection{Real time dynamics}

Once the structure of the isolated  droplet  has been obtained, it is used to prepare the initial 
state for   the merging. In all cases, we place two droplets on the $x$ axis 
such that their center-of-mass are at a  $d= 41$ \AA{} distance. We recall that the droplet sharp density surface, 
defined by a sphere of radius $R=r_0 N^{1/3}$ with $r_0=2.22$ \AA{},\cite{Bar06} is $R=17.6$ \AA{} for a $N=500$ droplet,
so that the  sharp density surfaces  of the vortex-free droplets are about 6 \AA{} apart. 

In classical droplet-merging calculations, the initial condition consists of
two droplets whose sharp-surfaces are touching, connected by tiny  bridge  
joining them.\cite{Egg99} In our case, the bridge appears naturally, caused by  the VdW mutual attraction. 
This attraction is absent in any GP approach based on an atom-atom  contact interaction, hence some initial velocity  
has to be provided to the droplets  or, alternatively,  
the confining potential has to be modified in order to trigger the process.\cite{Fer19,Sun08}
 
In our case, the combination 
\begin{equation}
\Psi(\mathbf{r},t=0)  = \Psi_L(\mathbf{r}-\frac{d}{2}\,\hat{{\bf x}}) + \Psi_R(\mathbf{r}+\frac{d}{2}\,\hat{{\bf x}}) \; ,
\label{eq33}
\end{equation}
is taken as initial configuration for the dynamics,
and the left ($\Psi_L$) and right ($\Psi_R$)  components may host vortices with given orientations.
The TDDFT dynamics is governed by the equation
\begin{equation}
i\hbar \frac{\partial}{\partial t} \Psi(\mathbf{r},t)  = \left[-\frac{\hbar^2}{2m}\nabla^2+\frac{\delta \mathcal{E}_c}{\delta \rho} 
 \right] \Psi(\mathbf{r},t)
 \label{eq4}
\end{equation}
which is solved as outlined in Refs. \onlinecite{Anc17,dft-guide} using the computing package of Ref. \onlinecite{Pi17}.
Most simulations span 600 ps of total elapsed time, with a time step of 1 fs.

We recall that if the effective wave function is written as
$$\Psi(\mathbf{r}, t)= \Phi(\mathbf{r}, t) \exp[\imath \,{\cal S}(\mathbf{r}, t)]$$ 
with $\Phi $ and ${\cal S}$ real functions, then the particle current density is
\begin{eqnarray}
{\mathbf j}({\mathbf r},t ) &=& - \frac{ \imath \; \hbar}{2 m} [\Psi^*({\mathbf r},t) \nabla \Psi({\mathbf r},t) - \Psi({\mathbf r},t) \nabla \Psi^*({\mathbf r},t)] 
\nonumber
\\
&&= \rho({\mathbf r},t)  {\mathbf v}({\mathbf r},t)
\end{eqnarray}
with $\rho ({\mathbf r},t)$ = $\Phi^2({\mathbf r},t)$
and ${\mathbf v}({\mathbf r},t) =   \hbar \, \nabla {\cal S}({\mathbf r},t)/m$; 
the phase ${\cal S}({\mathbf r},t)$ is thus the velocity potential field.

\begin{figure}[!]
\centerline{\includegraphics[width=1.0\linewidth,clip]{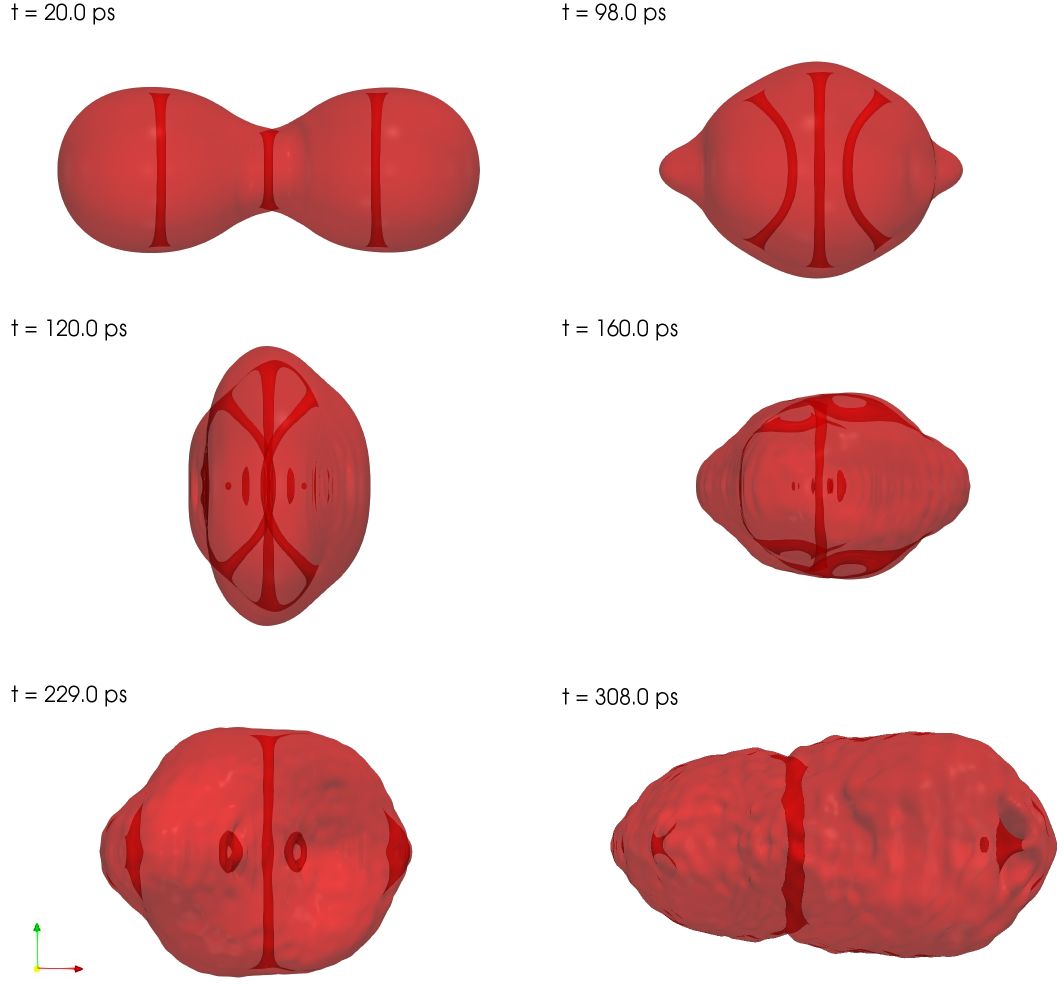}}
\caption{
Merging of two $^4$He$_{500}$ droplets, each initially hosting a vortex line perpendicular to the merging direction
with equal circulation number (vortex dimer, $\langle L \rangle = 1000 \hbar$). The 3D frames are labeled by the time in picoseconds.
}
\label{fig2}
\end{figure}
\begin{figure}[!]
\centerline{\includegraphics[width=1.0\linewidth,clip]{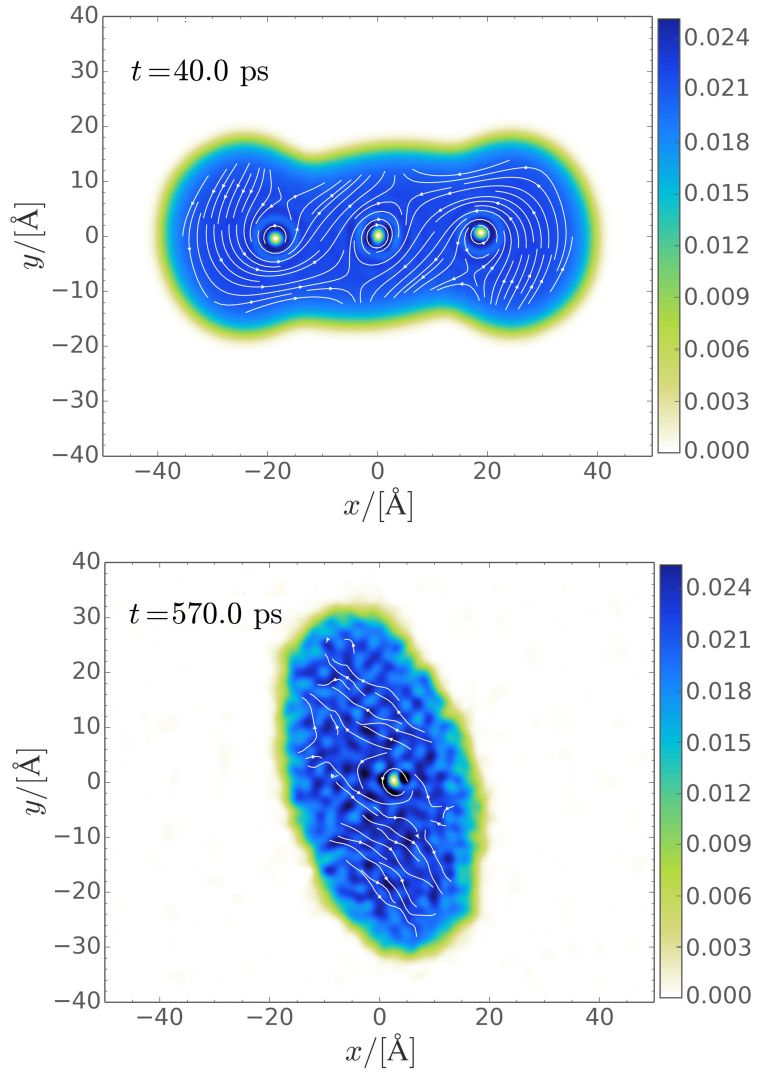}}
\caption{
Snapshots of the 2D density in the $x-y$ plane for the merging of two 
$^4$He$_{500}$ droplets, each initially hosting a vortex line along the $z$ axis
with the same circulation number (vortex dimer). Top:  $t=40$ ps, displaying the
cross section of the two initial vortex, and that of the nucleated antivortex between them. Bottom: same as top at $t=570$ ps.
Superposed to the density are circulation lines of the superfluid flow. The color bar indicates the value of the helium atom density in \AA$^{-3}$.
}
\label{fig3}
\end{figure}
\begin{figure}[!]
\centerline{\includegraphics[width=1.0\linewidth,clip]{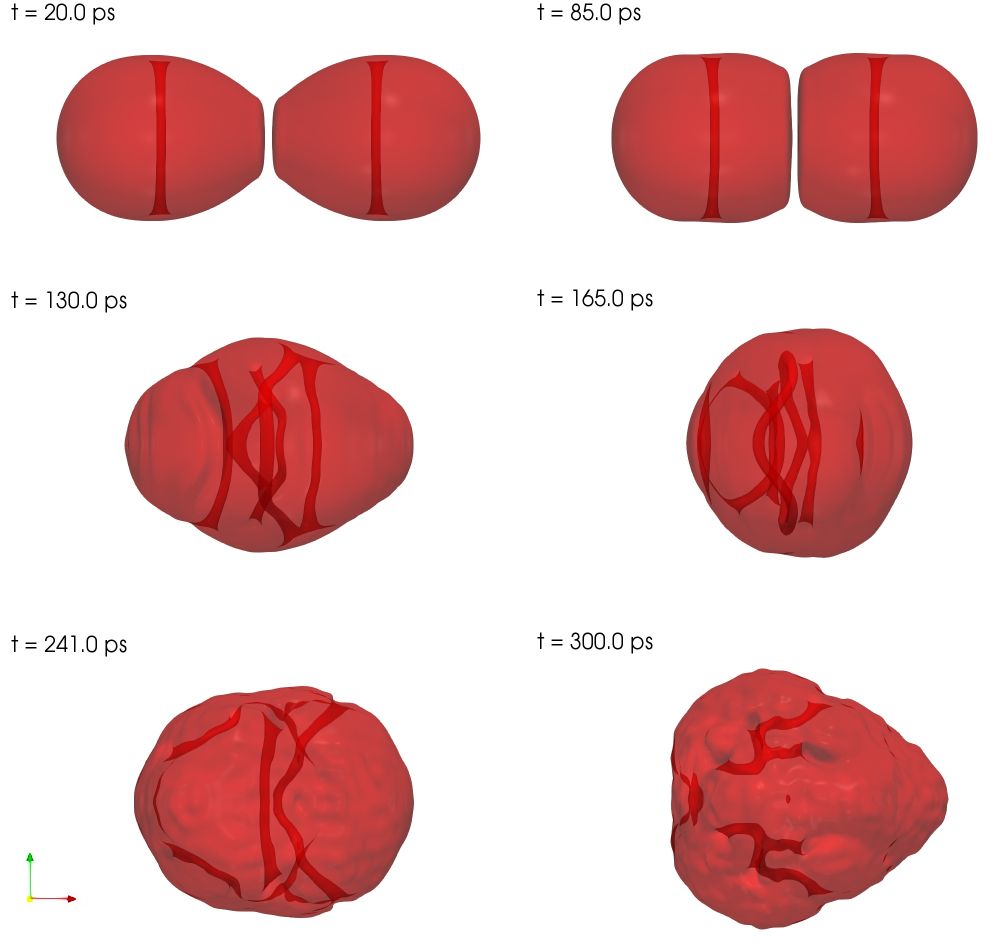}}
\caption{
Merging of two $^4$He$_{500}$ droplets, each initially hosting a vortex line along the $z$ axis
with opposite circulation number (vortex dipole, $\langle L \rangle = 0$). The 3D frames are labeled by the time in picoseconds.
}
\label{fig4}
\end{figure}

\section{Results}

\subsection{Merging dynamics}

In the vortex-free case, coalescence  smoothly proceeds by
the development of a tiny, low density bridge connecting the 
two droplets in about  4-5 ps, which eventually becomes a high density neck.\cite{Esc19}  The time evolution of the contact  region is  
dramatically affected by the presence  of vortices in the droplets and the interplay between their associated flows.\cite{SM}

Figure \ref{fig2} shows snapshots of the merging droplets  in the case of two parallel vortices, one in each droplet, 
 perpendicular to the merging direction and with the same quantum circulation
(vortex dimer, see Fig. \ref{fig1}), which has a total angular momentum $\langle L \rangle = 1000 \hbar$.  
The three-dimensional (3D) droplets are represented by drawing their sharp density 
surfaces, defined as the loci where  the helium density equals that of bulk liquid at zero 
temperature and pressure divided by two, namely  0.0109 \AA$^{-3}$. 

At $t=20$ ps, a vortex line parallel to the original ones appears in the neck region connecting the  
merging droplets; since angular momentum is strictly conserved,  its initial value is 
shared between the original vortices, the nucleated one and the surface capillary waves which appear in the merged droplet.
Intuitively, the vortex nucleated at the neck should have a circulation opposite to the original ones.\cite{Sun08} 
We show in Fig. \ref{fig3} the circulation lines of the superfluid flow at $t=40$ ps in the $x-y$ plane.

At $t=98$ ps, the original vortex lines have experienced a large bending in order to hit perpendicularly the droplet surface. 
As in the vortex-free case, two protrusions appear on the droplet surface, whose subsequent collapse 
in the body of the droplet launches a series of vortex rings.
The appearance of surface protrusions is frequent in classical droplets collisions.\cite{Pan09,Men01,Pan05,Hua19}
What makes protrusions unique in the case of $^4$He droplets is that they act as nucleation sites of quantized vortex rings.
Surface protrusions have also been identified in  head-on collisions of self-bound quantum droplets,\cite{Fer19,Cik21} but
no vortex ring nucleation has been reported in these studies. For these droplets made of ultradilute Bose-gase mixtures, a minimum 
number of atoms is needed to develop any binding, and the surface tension is very small. 
As a result, once created the 
protrusions evaporate almost immediately instead of being re-adsorbed within 
the merging droplet, hindering the nucleation of vortex rings. 
The helium-drop mechanism should also work for colliding classical droplets, but we are not aware that 
non-quantized vortex ring nucleations have been reported in droplets collisions.

The frame corresponding to $t=120$ ps shows the appearance of a vortex tangle and of vortex rings as well. 
The
disentanglement of the vortex lines following their mutual crossing yields vortex loops, as shown e.g. at $t=160$ ps. The dynamic 
evolution also yields vortex reconnections and the nucleation of vortex rings, as displayed at $t=229$ ps. These vortex rings eventually
collide with the droplet surface and evanesce.\cite{SM}  Finally, 
a single vortex line emerges, as displayed at $t=308$ ps,  which remains till the end of the simulation while the droplet is still in a highly excited state. 
The two-dimensional (2D) density on the $x-y$ plane corresponding to the  configuration at $t=570$ ps is shown in the bottom panel of Fig. \ref{fig3}.

\begin{figure}[!]
\centerline{\includegraphics[width=1.0\linewidth,clip]{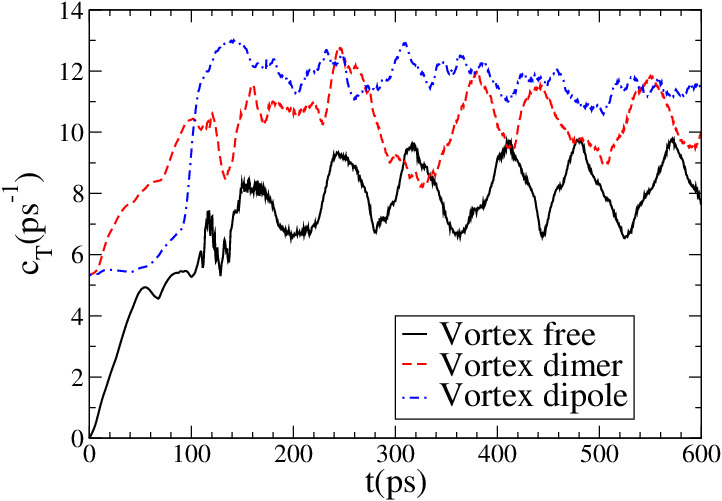}}
\caption{ 
Space integral of  $\|\boldsymbol{\upomega}_{ps}(\mathbf{r}, t)\|$  ($c_T(t)$ function) 
as a function of time for the merging of two $^4$He$_{500}$ droplets.
}
\label{fig5}
\end{figure}

It is worth noticing the presence of helical Kelvin modes along the vortex core, that can only appear if
it has some transverse extension. Helical Kelvin modes have also been found in the capture of impurities
by vortex lines in helium droplets.\cite{Cop17} Superposed to this complex vortex dynamics arising from surface deformations,
the merged droplet undergoes large amplitude oscillations.\cite{Esc19} 

Figure \ref{fig3} shows that, besides circulation lines that wrap around the vortex cores, there are others that hit
the droplet surface and correspond to surface capillary waves. Surface capillary waves are ubiquitous in deformed superfluid droplets 
and carry a part --sometimes sizeable-- of the angular momentum of the droplet.\cite{Anc18,Oco20,Pi21}

\begin{figure}[hbt!]
\centerline{\includegraphics[width=0.9\linewidth,clip]{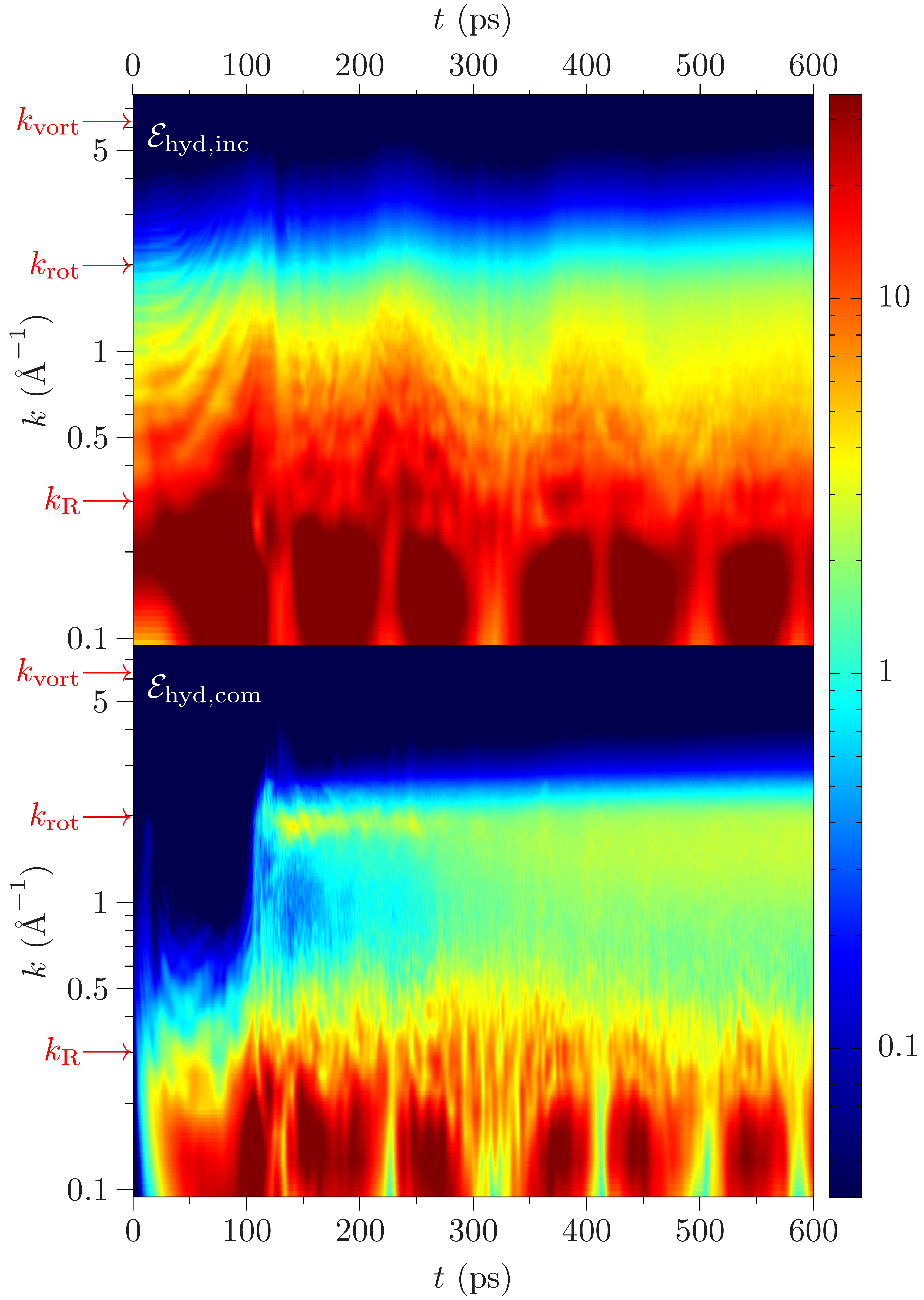}}
\caption{
Energy spectrum $\mathcal{E}_{\textrm{hyd}}(k,t)$ (K\,\AA)
for the vortex dimer. Top: incompressible part; bottom: compressible part.}
\label{fig6}
\end{figure}

\begin{figure}[hbt!]
\centerline{\includegraphics[width=0.9\linewidth,clip]{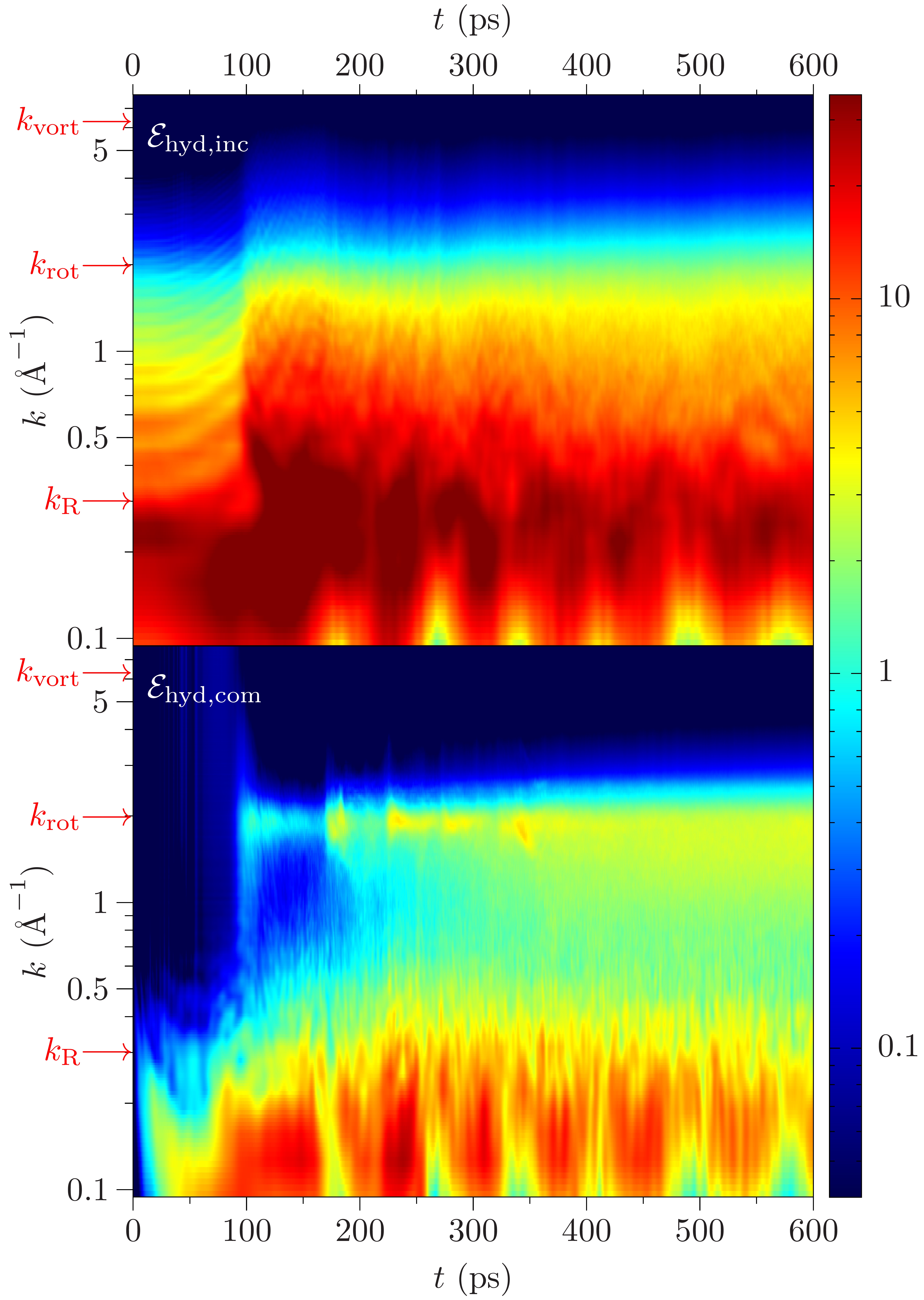}}
\caption{
Energy spectrum $\mathcal{E}_{\textrm{hyd}}(k,t)$ (K\,\AA)
for the vortex dipole. Top: incompressible part; bottom: compressible part.}
\label{fig7}
\end{figure}

We have next addressed the coalescence of droplets  in the case of two parallel vortices, one in each droplet, perpendicular to the merging direction and
with opposite quantum circulation (vortex dipole, see Fig. \ref{fig1}), which has a total angular momentum $\langle L \rangle = 0$. 
Since the phase difference between the superfluid 
droplets at the touching point is $\pi$, a dark soliton is expected to appear at the contact region.
We recall that a dark soliton is a structure (a disk in the present case) made  
of nodal points of the effective wave function $\Psi(\mathbf{r},t)$ were the real and imaginary parts are both zero;\cite{Pit16}
this is at variance with vortices, where the nodal points form a line.

Bright solitons in thin $^4$He films adsorbed on solid substrates have been observed,\cite{Kon81,McK90,Hop96,Lev97}
and also in bulk superfluid $^4$He;\cite{Anc18b,Bue19} so far, dark solitons have not been detected in superfluid $^4$He. Instead, they have been observed in
BECs.\cite{Bur99,Den00,And01,Bec08} While a soliton in 1D is stable, the presence of a transverse dimension 
causes its decay through the snake instability mechanism.\cite{Mur99,Fed00} In Ref. \onlinecite{Ku16} the time evolution and decay of a dark
soliton was observed. The cascade of solitonic excitations arising from the snaking of a planar soliton in BEC has recently been 
observed.\cite{Don14}

Figure \ref{fig4} shows snapshots of the merging of the dipole configuration and  the appearance of a dark soliton, fully developed at $t=85$ ps. 
The droplets are prevented from merging for about 130 ps  because the phase difference makes the touching plane a nodal surface.
The dark soliton in Fig. \ref{fig4} is found to undergo a snake instability, decaying into vortex rings 
that in turn decay into a vortex-antivortex tangle which  can be observed in the $t=130$ ps frame.\cite{SM}  
Upon vortex reconnections, a distorted vortex ring appears at $t=165$ ps.
Until the end of the simulation, vorticity appears in the bulk of the excited merged droplet, which is very deformed and has many bumpy textures on its surface,
as seen e.g. in the frame at $t=300$ ps. It can be seen that in the dipole case the merging yields 
many vortex reconnections;\cite{SM} at variance,  in the dimer case the vortex that appears as soon as 
the droplets touch remains stable for a rather long period of time.

A dark soliton similar to the one described before can be created by starting 
the real time evolution from the effective wave function 
$\Psi_0(\mathbf{r}-d/2\,\hat{{\bf x}}) + e^{\imath \pi} \,
\Psi_0(\mathbf{r}+d/2\,\hat{{\bf x}})$, $\Psi_0$ being the vortex-free 
effective wave function (see Fig. \ref{fig1}).\cite{SM} A similar preparation for two colliding confined BECs was made in Ref. \onlinecite{Sun08}. 
As in the vortex dipole case just described, we have found that
the dark soliton decays by vortex rings nucleation, which in turn decay into a vortex tangle.

When vortices  are oriented perpendicularly to each other and to the merging direction as well (see Fig. \ref{fig1}), droplet coalescence 
 yields a vortex line at the contact region forming an angle of $\pi/4$ with respect to the original ones. 
The nucleated vortex is less robust than that appearing in the dimer case, 
which is aligned with the original ones  for quite some time.\cite{SM} 
For this reason, although the time evolution of the merging process is  rather similar for both geometries, we have found that 
vortices become more entangled and yield more reconnections and splittings in the perpendicular than in the dimer case.

We have also addressed the merging of  droplets with vortices  oriented in other directions.\cite{SM} In particular,
configurations where two vortices are aligned with the  merging direction, either with the same  circulation  
(aligned vortices, see Fig. \ref{fig1}), or with opposed circulation (anti-aligned vortices, see Fig. \ref{fig1}). 

In the aligned case, the simulation has unveiled an unusual scenario.
Whereas the final state is expectable --a single vortex line arising from the connection 
of the two aligned vortices-- a series of vortex-antivortex ring  pairs are dynamically nucleated at the surface protrusions as in the vortex-free 
merging.\cite{Esc19} 
These rings are pierced by the vortex line, which prevents them from shrinking, and guides them
till they collide and annihilate at the center of the merged droplet producing a roton burst. 

The anti-aligned case has also yielded its own peculiarities.\cite{SM}  The droplet coalescence produces two new vortex lines perpendicular
to the merging direction, forming a cross patt\'ee (similar to the Maltese cross) with opposite circulations such that the total angular 
momentum in the merged droplet is kept equal to zero.  The collapse of the surface protrusions yields the usual vortex-antivortex ring pairs, 
pierced again by the anti-aligned vortex lines. 
The annihilation of the vortex rings produces helical Kelvin modes along the vortex lines which have been nucleated perpendicularly to the merging
direction, clearly visible before the cross structure breaks down into a vortex-antivortex pair whose constituents reconnect, forming a large vortex ring 
nearly occupying the cross section of the merged droplet before breaking down again into a vortex-antivortex pair, and so on. These splittings and 
reconnections produce roton bursts of different intensities.\cite{SM} 

 Finally, we have studied the coalescence of a vortex-free droplet with a droplet hosting a vortex line aligned with the merging direction.
 This simulation has been motivated by a recent work,\cite{Kan20}  where two laterally confined cylindrical cold-gas BECs  
 are  kept apart by a planar potential barrier between them. One of the BECs is  vortex-free and the other hosts a vortex
 line along the symmetry axis. Suddenly, they are put in contact by releasing the barrier, merging in the direction of the vortex line. 
 A striking result found in this work, is that  a ``corkscrew'' structure emerges at the BECs interface by which 
angular momentum is transferred from the vortex-hosting to the vortex-free BEC: the vortex line deforms adopting the shape of a corkscrew, penetrating into the vortex-free BEC. 

In the merging of superfluid He droplets, the contact region is not planar and droplets deform laterally, so 
it is interesting to see how the scenario changes. 
For both systems, the evolution of the vortex line during the first stages of the simulation  is similar:\cite{SM} 
since a vortex line cannot end in the bulk of the superfluid,  it bends off the merging axis in order to hit perpendicularly the droplet surface.
The subsequent evolution is however very different, 
as the surface  of the $^4$He droplet is dramatically deformed by the merging process and the behavior of the 
vortex line is erratic, whereas in the confined BECs its time evolution is guided by the lateral confinement, to which the vortex line
has to hit perpendicularly.  Eventually, both dynamics become similar again, with the original vortex line cut into several vortex loops
and the appearance of a very distorted superfluid density. 

\subsection{Vorticity and energy spectrum}

\begin{figure}[hbt!]
\centerline{\includegraphics[width=1.0\linewidth,clip]{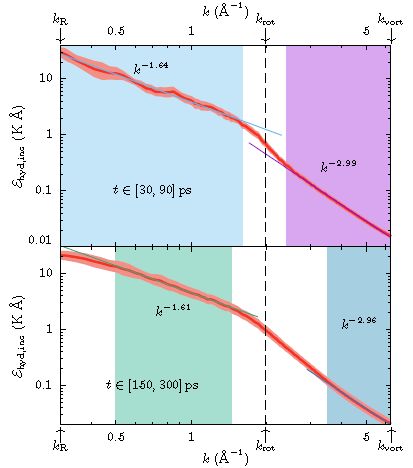}}
\caption{
Power-law analysis of time averages of the incompressible energy spectrum for the  dimer configuration. 
Each of the two panels display, in log-log scales, the time averaged spectrum ${\cal E}(k)$ (red line) and the range
within a standard deviation (light red) for every $k$, as well as the power laws determined by weighted fits over the shadowed
$k$ intervals and indicated time intervals. Top panel: $k^{-1.64}$  and $k^{-2.99}$ for $t \in [30, 90] ps$. Bottom panel:
$k^{-1.61}$  and $k^{-2.96}$ for $t \in [150, 300] ps$
}
\label{fig8}
\end{figure}

\begin{figure}[hbt!]
\centerline{\includegraphics[width=1.0\linewidth,clip]{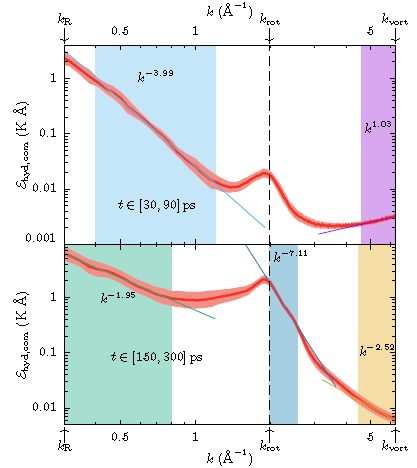}}
\caption{
Power-law analysis of time averages of the compressible energy spectrum for the  dimer configuration. 
Each of the two panels display, in log-log scales, the time averaged spectrum ${\cal E}(k)$ (red line) and the range
within a standard deviation (light red) for every $k$, as well as the power laws determined by weighted fits over the shadowed
$k$ intervals and indicated time intervals. Top panel: $k^{-3.99}$  and $k^{1.03}$ for $t \in [30, 90] ps$. Bottom panel:
$k^{-1.95}$, $k^{-7.11}$  and $k^{-2.52}$ for $t \in [150, 300] ps$
}
\label{fig9}
\end{figure}

Detecting vorticity, defined as $\boldsymbol{\upomega}(\mathbf{r}, t)= \nabla \times {\mathbf v}(\mathbf{r}, t)$,\cite{Don91}
is not a trivial task, since it is singular  on the vortex lines  and zero outside them (a Dirac delta function).\cite{Pit16}
A more convenient way to identify regions of potentially non-zero vorticity in the superfluid
is to calculate the curl of the atom current density, 
$\boldsymbol{\upomega}_{ps}(\mathbf{r}, t)= \nabla \times {\mathbf j}(\mathbf{r}, t)$, called pseudo-vorticity.\cite{Vil16}
Plots of the iso-surfaces $\|\boldsymbol{\upomega}_{ps}(\mathbf{r}, t)\|= C$ 
allow to visualize regions of non-zero pseudo-vorticity,\cite{Esc19} that may be subsequently analyzed seeking for 
vorticity inside them.

Similarly, the integrated quantity  
\begin{equation}
c_T(t) = \int {\it d} {\mathbf r} \,  \|\boldsymbol{\upomega}_{ps}(\mathbf{r}, t)\|
\label{eq6}
\end{equation}
helps identify the appearance of pseudo-vorticity in time.
Figure \ref{fig5} shows the function $c_T(t)$  for the  vortex dimer and vortex dipole droplets merging. 
It starts at a common non-zero value,  as $c_T(t)$ does not depend on the relative orientation of the vortices. 
For the sake of comparison, the result corresponding to the vortex-free droplet merging is also shown. 
In this case, the steadily raising of $c_T(t)$ up to $t \sim 50$ ps is due to the nucleation of vortex-antivortex ring pairs 
at the contact region, that slip on the surface of the coalescent droplets.\cite{Esc19} It was show  that quantized ring vortices 
appeared on the surface of the merged droplet up to much later times, of the order of 115 ps.

In the dimer case, the observed steadily increase of  $c_T(t)$ up to $t \sim 100$ ps is mainly due to the nucleation of the antivortex line
in the merging region, with some contribution from the vortex-antivortex ring  pairs as in the vortex-free case. 
 In both cases, $c_T(t)$ has contributions from the  vortex rings nucleated at the surface protrusions.
 
In the dipole case,  Fig. \ref{fig5}  shows that vorticity is nearly constant
up to about $t=50$ ps, slowly increasing during the next 40 ps due to  vortex-antivortex ring pairs nucleated in the low density region 
close to the nodal plane. The sharp rise at about $t=90$ ps is produced by the decay of the dark soliton into 
quantized vortex rings, whose decay  yields vortex-antivortex pairs. 
From this time on, pseudo-vorticity remains fairly constant during the elapsed time of the simulation (600 ps).

\begin{figure}[hbt!]
\centerline{\includegraphics[width=1.0\linewidth,clip]{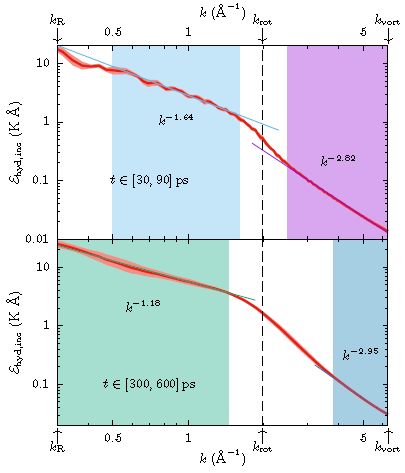}}
\caption{
Power-law analysis of time averages of the incompressible energy spectrum for the  dipole configuration. 
Each of the two panels display, in log-log scales, the time averaged spectrum ${\cal E}(k)$ (red line) and the range
within a standard deviation (light red) for every $k$, as well as the power laws determined by weighted fits over the shadowed
$k$ intervals and indicated time intervals. Top panel: $k^{-1.64}$  and $k^{-2.82}$ for $t \in [30, 90] ps$. Bottom panel:
$k^{-1.18}$  and $k^{-2.95}$ for $t \in [300, 600] ps$
}
\label{fig10}
\end{figure}

\begin{figure}[hbt!]
\centerline{\includegraphics[width=1.0\linewidth,clip]{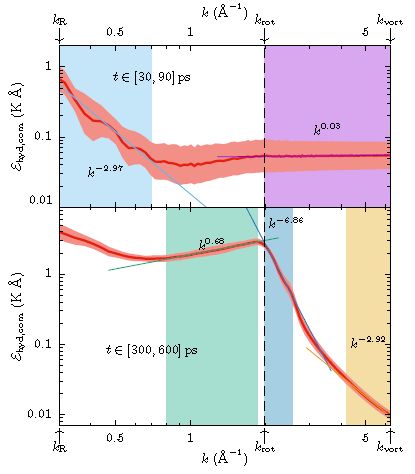}}
\caption{
Power-law analysis of time averages of the compressible energy spectrum for the  dipole configuration. 
Each of the two panels display, in log-log scales, the time averaged spectrum ${\cal E}(k)$ (red line) and the range
within a standard deviation (light red) for every $k$, as well as the power laws determined by weighted fits over the shadowed
$k$ intervals and indicated time intervals. Top panel: $k^{-2.97}$  and $k^{0.03}$ for $t \in [30, 90] ps$. Bottom panel:
$k^{0.68}$, $k^{-6.86}$  and $k^{-2.92}$ for $t \in [300, 600] ps$
}
\label{fig11}
\end{figure}

Figure \ref{fig5} and a similar one in the Supplemental Material\cite{SM} for the aligned and anti-aligned vortex orientations,
displays conspicuous vorticity oscillations.
From the analysis of the real-time evolution of the merging process\cite{SM} 
it appears that the configurations associated with the maxima in $c_T(t)$  
correspond to the more compact, more axisymmetric-like
configurations of the merged droplet, while the minima correspond 
to the more elongated, less axisymmetric-like configurations. 
Therefore we attribute tentatively the oscillations in $c_T(t)$ 
to the interplay between vortex lines and capillary 
waves: 
while they share the angular momentum 
deposited in the merged droplet, only vortex lines carry vorticity. Since capillary waves only appear in  non-axisymmetric configurations,  
vortices should be more abundant in compact configurations to conserve the angular momentum, hence increasing 
the $c_T(t)$ value. This reasoning cannot be applied to the other cases.

It is worth noticing that the surface of the merged droplet evolves from smooth to rough. This roughness is remarkably similar to
that found on the surface of classical fluids --ethanol and water-- in low gravity,\cite{Fal09} and it is a hallmark of capillary (classical) 
wave turbulence.\cite{Naz15}
Indeed, nucleation of vortex lines and rings and their reconnections, as well as their subsequent annihilation  
during the merging, are expected to be a source of quantum turbulence. Vortex ring emission and the possible transition to
chaotic turbulent regime occurring when impurities move inside a superfluid  has been  previously studied  within the  time-dependent GP approach
in both superfluid $^4$He and   BECs.\cite{Bar01,Vil18,Ber01,Fri92,Nav16,Ber02,Kob05,Pro09}
Although the GP equation is too simple a  model for superfluid helium, it has been shown
to reproduce the dynamics of vortex ring emission  for an electron moving in the liquid at zero temperature, as
well as the inherent symmetry breaking of the solution due to the emerging instability, see e.g. Ref. \onlinecite{Vil18} and references therein. 

Turbulence  is usually addressed with the aid of the kinetic energy spectrum whose dependence upon the wavenumber $k$ allows
to distinguish different regimes that are relevant for characterizing it.\cite{Tsu09} The kinetic energy of the superfluid is written as
\begin{equation}
E_{\textrm{kin}}(t) =  \frac{\hbar^2}{2m}  \int \dd {\mathbf r} \, \left|\nabla \sqrt{\rho({\mathbf r},t)}\right|^2
+ \frac{m}{2} \int \dd {\mathbf r} \,  \frac{{\mathbf j}^2({\mathbf r},t)}{\rho({\mathbf r},t)}
\; ,
\label{A1bis}
\end{equation}
where the first term is the quantum pressure while the second is the usual hydrodynamic kinetic energy $E_{\textrm{hyd}}$.
In Fourier space,  $E_{\textrm{hyd}}$  is rewritten as
\begin{equation}
E_{\textrm{hyd}}(t) =  4 \pi \int_0^\infty  \dd{k}  \, \mathcal{E}_{\textrm{hyd}}(k,t) \; ,
\label{eq9}
\end{equation}
where the energy spectrum $\mathcal{E}_{\textrm{hyd}}(k,t)$ is the
spherical average in $\mathbf{k}$-space\cite{Bar01,Nor97,Bra12}
\begin{equation}
{\mathcal{E}}_{\textrm{hyd}}(k,t) = \frac{m}{2} (2 \pi)^3  \frac{k^2}{4\pi}
\int_{\Omega_k}  \dd\Omega_k
\left| \widetilde{\frac{ {\mathbf j}}{\sqrt{\rho}}}({\mathbf k},t)\right|^2
\;.
\label{eq10}
\end{equation}
$\mathcal{E}_{\textrm{hyd}}(k,t)$ is decomposed it into a divergence-free (incompressible) part, related to vorticity,
and a compressible part,\cite{Nor97} related to density waves, which are  analyzed separately.\cite{Ree12}
$k$ values relevant for the discussion are $k_{\textrm{vort}}= 2\pi/a = 6.3$\,\AA$^{-1}$, $a$ being the vortex core radius $\sim 1$\,\AA{};
$k_{\textrm{R}}=2\pi/R= 0.3$\,\AA$^{-1}$, where $R$ is the radial dimension of the droplet, and
the roton wave-vector $k_{\textrm{rot}}=1.98$\,\AA$^{-1}$.
Figure \ref{fig6} shows both additive components of $\mathcal{E}_{\textrm{hyd}}(k,t)$
for the vortex dimer, which periodically display bright regions in the small $k$ region ($k\lesssim k_{\mathrm{R}}$)
in phase with the oscillations of the shape of the merged droplet. Fig. \ref{fig7} shows the results for the vortex dipole.

In the compressible component (bottom panel in both figures), a series of bright spots
appears arising from rotons emission.\cite{Esc19}
The simultaneous analysis of the droplet densities\cite{SM} and the energy spectrum shows that 
roton bursts originate from  three mechanisms, namely the annihilation of  vortex-antivortex rings, which yields the more intense peaks,
vortex-antivortex reconnections, and vortex ring annihilation when they hit the droplet surface.
The peaks spread with time, being clearly visible  during  several hundred picoseconds more.
It is worth noticing that in a recent paper, the head-on collision of a pair of smoke vortex-antivortex rings has been visualized,\cite{McK18}
as well as  their ultimate breakdown yielding a burst of sound waves similar to that we have found for rotons, stressing
its connection to the appearance of classical turbulence. 

As in Ref. \onlinecite{Esc19}, we have carried out power-law ($k^{\alpha}$) analyses for selected  time average intervals
 of the compressible and incompressible 
spectra for the droplet mergings addressed in this work. Whereas it is not possible to draw a common and solid explanation
for all the observed values of $\alpha$,  values close to $\alpha =-5/3$ and $-3$ often appear in the incompressible spectrum, which  are 
respectively associated  to the appearance of  
classical (Kolmogorov) and superfluid turbulence 
in the bulk,\cite{Nor97,Sal10} and to the presence of vortices.\cite{Nor97,Bra12}

Figures \ref{fig8} and  \ref{fig9} show the power-law analysis for the dimer configuration.
The first time interval (top panels in the figures, 30 to 90 ps) corresponds to the dynamics in the presence of three vortices: it starts
after the central vortex is well formed, and finishes before the protrusions of the merged droplet (already largely depleted) accelerate 
towards the center of the droplet. The second time interval (bottom panels, 150 to 300 ps) addresses a later 
time interval with intricate vortex
dynamics after the roton burst. The incompressible spectrum remains quite constant over this time period, but the compressible spectrum
changes noticeably over the interval, as the roton peak becomes less prominent as times passes.

Finally, Figs. \ref{fig10} and  \ref{fig11} show the power-law analysis for the dipole configuration.
The first time interval (top panels in the figures, 30 to 90 ps) corresponds to the dynamics where the dark soliton develops, before the droplets
merge. It is worth seeing the $\sim k^{-5/3}$ law found in the incompressible spectrum. In the compressible spectrum we have found a progressive
widening and strengthening of the roton peak and the possibility of fitting in all cases power laws with exponents close to $-3$ for large $k$ values. Let
us recall that the multimedia materials accompanying this work \cite{SM} allows one to dynamically follow the appearance of the compressible and 
incompressible spectra as a function of time.

\section{Summary}

In this work we have simulated the coalescence of superfluid 
helium droplets  hosting vortex lines in different relative orientations, finding a rich  
phenomenology, as {\it e.g.} the appearance of vortex rings slipping along vortex lines and the appearance  of dark solitons from the annihilation of 
vortex-antivortex pairs and their decay into vortex rings. 

The merging process generates vorticity which adds up to the pre-existing one.  In the present study, it builds up
 in about 100 ps (the time to produce the merged droplet) and it is as large as that originally  in the droplets. 
 The generated vorticity is fairly the same irrespective of the relative orientation of the vortex lines in the merging droplets, indicating that
 its origin has more to do with the droplet dynamics than with the relative orientation of the vortices. 
This vorticity appears as a genuine surface effect, being especially apparent in the vortex-free case where no vorticity is found 
in the bulk of the merging droplets during the first 100 ps.

We cannot ascertain from our time-limited simulations whether (and how) vorticity fades away.
Part of this vorticity disappears  by atom evaporation; however,  one
should have in mind that  in spite of being highly excited states of the system, the simultaneous constraints of energy 
and angular momentum conservation
would prevent the decay of vortex lines in a helium droplet.\cite{Leh03} It might thus appear that vortex lines in helium droplets
are rather stable. As a matter of fact, they have been detected in large droplets on the millisecond timescale.\cite{Gom14,Oco20}

With the experimental  realization of stable self-bound ultradilute ``quantum droplets'' made of bosonic atoms
of a binary mixture,\cite{Cab18}
which  have been found to sustain quantum vortices,\cite{Anc18c}
it has been possible to study the collision of self-bound BECs.\cite{Fer19,Cik21} 
Thus, it should also be possible to  study  the merging of quantum droplets and
the influence of surface effects in the transfer of angular momentum from one quantum droplet to another, 
as well as a deeper analysis of the role of the surface protrusions   as potential nucleation sites of vortex rings, thus
connecting the ultradilute quantum droplet BECs and the superfluid $^4$He droplets scenarios.
 
\begin{acknowledgments}
This work has been  performed under Grant No.  PID2020-114626GB-I00 from the MICIN/AEI/10.13039/501100011033.
J.M.E. acknowledges support from the Spanish Research Agency (AEI)
through the Severo Ochoa Centres of Excellence programme (grant SEV-2017-0706)
and from the European Union MaX Center of Excellence (EU-H2020 Grant No.\ 824143).
\end{acknowledgments}

\newpage

\section{\bf Supplemental Material}

In this document we complement the results presented in the paper showing those corresponding to
the merging of two $^4$He$_{500}$ droplets hosting each a vortex line
when the vortices are aligned with the merging direction 
and  have the same quantum circulation  (aligned configuration), and when they have opposed circulations
(antialigned configuration), as well as the merging of a vortex-free $^4$He$_{500}$ droplet with a  $^4$He$_{500}$ droplet
hosting a vortex line along the merging direction  (``corkscrew'' configuration). 

We  have represented the 3D droplet densities  by their sharp 
surfaces, i.e., the loci at which the  helium density equals $\rho_0/2$, $\rho _0$ being the bulk liquid density.
We recall that the center-of-mass of the merging droplets are initially 41 \AA {} apart. Since the  presence of vortices in 
the droplets flatten them out, depending on the vortex orientation their surfaces are at slightly different distances and the Van der Waals attraction
acts with different intensity. This introduces some delay in the merging process which depends on the initial arrangement.

\begin{figure*}[!]
\centerline{\includegraphics[width=1.0\linewidth,clip]{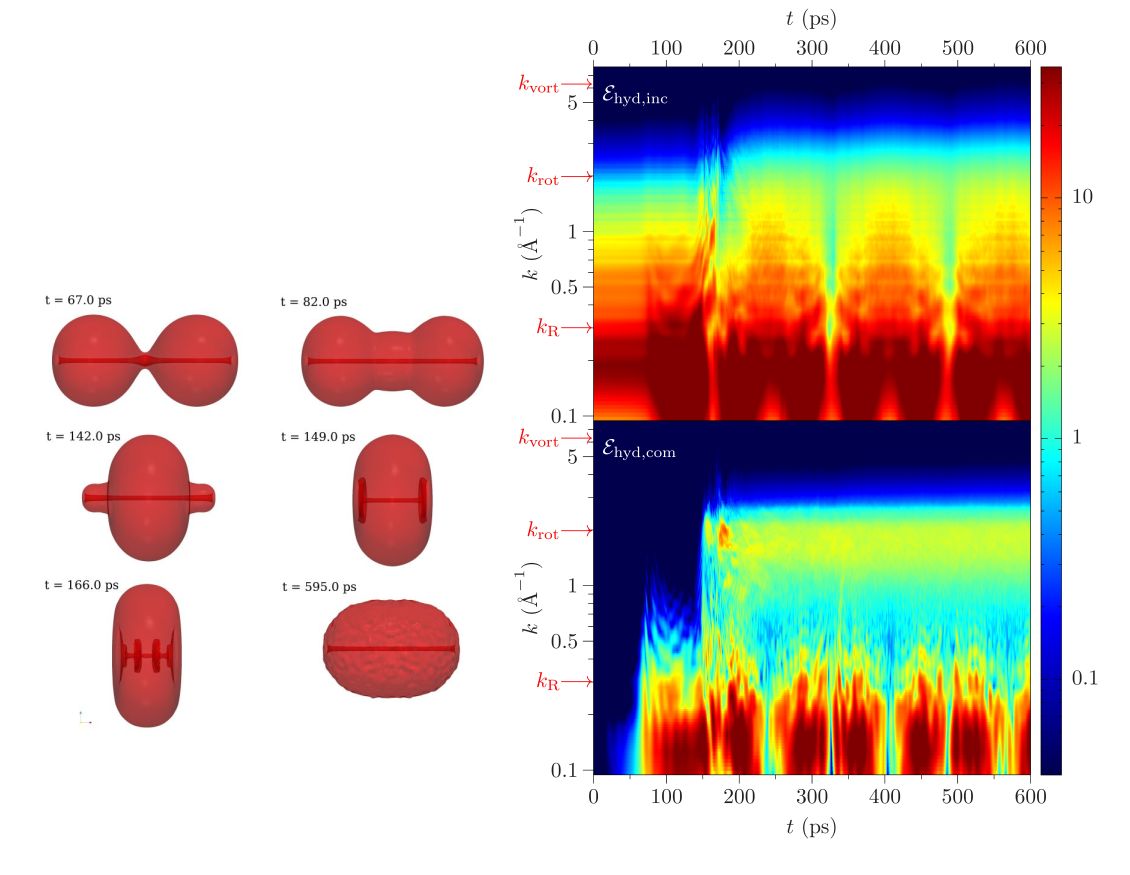}}
\caption{Merging of two $^4$He$_{500}$ droplets   
hosting each a vortex line  when the vortices are aligned with the merging direction 
and  have the same quantum circulation (aligned configuration).
Left: Snapshots of  the 3D density  at the labeled times.
Right: Energy spectrum $\mathcal{E}_{\textrm{hyd}}(k,t)$ (K\,\AA).Top, incompressible part; bottom, compressible part.
}
\label{fig1-SM}
\end{figure*}

\begin{figure*}
\centerline{\includegraphics[width=0.7\linewidth,clip]{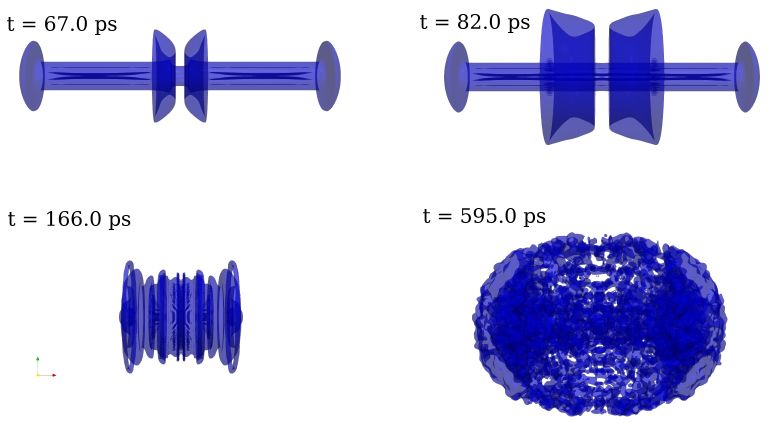}}
\caption{
Pseudovorticity $\boldsymbol{\upomega}_{ps}(\mathbf{r}, t)= \nabla \times {\mathbf j}(\mathbf{r}, t)$,
 visualized by $\|\boldsymbol{\upomega}_{ps}(\mathbf{r}, t)\|= C$ isosurfaces for $C=10^{-3}$ ps$^{-1}$
 at the labeled times, for the merging of two $^4$He$_{500}$ droplets   
hosting each a vortex line  when the vortices are aligned with the merging direction 
and  have the same quantum circulation (aligned configuration). 
}
\label{fig2-SM}
\end{figure*}

\begin{figure*}[!]
\centerline{\includegraphics[width=0.7\linewidth,clip]{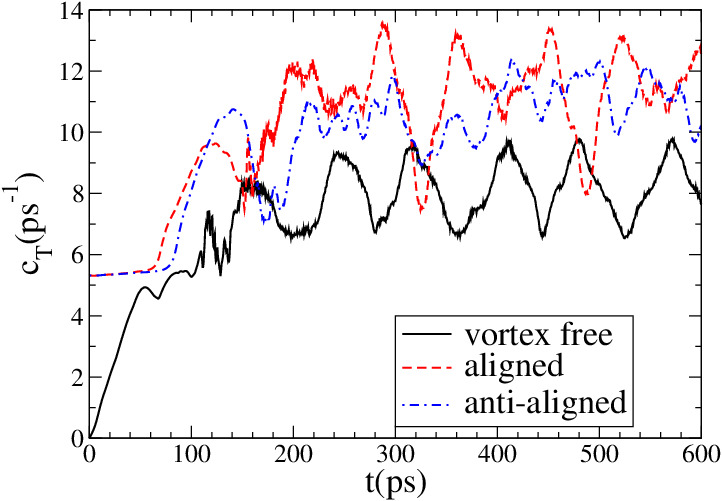}}
\caption{ 
Space integral of  $\|\boldsymbol{\upomega}_{ps}(\mathbf{r}, t)\|$  ($c_T(t)$ function) 
as a function of time for the merging of two $^4$He$_{500}$ droplets.
}
\label{fig3-SM}
\end{figure*}

Figure \ref{fig1-SM} left shows snapshots of the 3D density corresponding to the merging  of the aligned configuration.
The vortex lines fuse after a neck has developed connecting the
droplets. Simultaneously, quantized vortex-antivortex  ring pairs appear on the droplet surface 
as shown in Fig. \ref{fig2-SM}. These  rings are nucleated at the contact region 
and slip on the surface of the coalesced droplet during the merging process.\cite{Esc19} 
The associated vorticity yields the sharp rising of $c_T$ at $t\sim 60$ ps clearly visible on Fig. \ref{fig3-SM}.

The collapse of the two protrusions appearing symmetrically located  on the surface (snapshot at $t=142$ ps in Fig. \ref{fig1-SM}) 
yields two trains of vortex-antivortex  ring pairs that move against each other in the bulk of the droplet. 
 Remarkably, the vortex line arising from the connected linear vortices pierces these rings, preventing their
 shrinking and guiding them till they collide and annihilate at the center of the merged droplet. This produces an intense roton burst at
 $t =175$ ps, as shown in the bottom right panel of Fig. \ref{fig1-SM}. 
Eventually, a robust vortex line remains, together with a  bumpy droplet surface where wave turbulence sits\cite{Fal09,Naz15,Esc19}
 while the merged droplet undergoes large amplitude oscillations. 
 
\begin{figure*}
\centerline{\includegraphics[width=1.0\linewidth,clip]{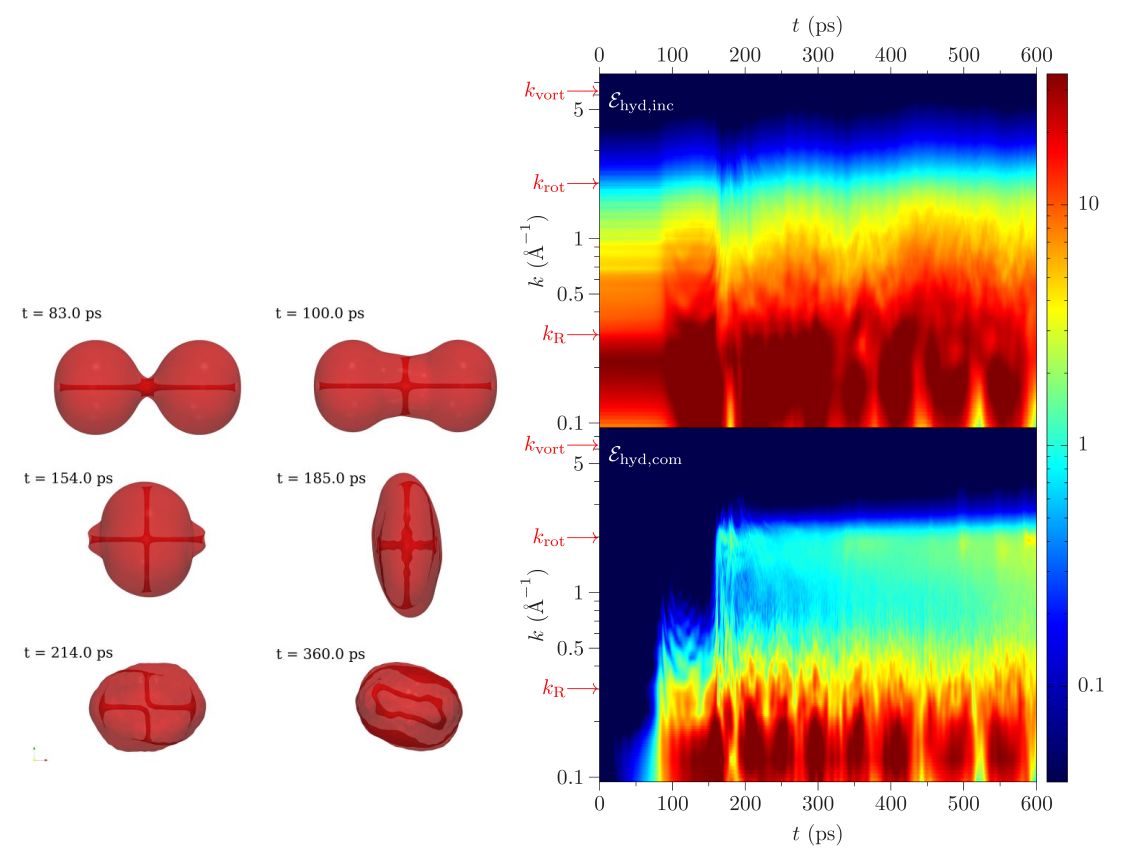}}
\caption{Merging of two $^4$He$_{500}$ droplets   
hosting each a vortex line  when the vortices are aligned with the merging direction 
and  have opposed quantum circulations (anti-aligned configuration).
Left: Snapshots of  the 3D density  at the labeled times.
Right: Energy spectrum $\mathcal{E}_{\textrm{hyd}}(k,t)$ (K\,\AA).Top, incompressible part; bottom, compressible part.
}
\label{fig4-SM}
\end{figure*}

Figure \ref{fig4-SM} left shows snapshots of the 3D density corresponding to the merging  of the anti-aligned configuration.
It is worth noticing the appearance at the contact region of  two vortex lines perpendicular to the direction of 
coalescence, each one having a different quantum circulation.
 This cross structure lasts for  quite a long period of time, eventually breaking into a vortex-antivortex pair 
(snapshot at $t=214$ ps).  It is also worth seeing the appearance of distinct helical Kelvin modes along the vortex lines 
 (snapshot at $t=185$ ps), and of a distorted vortex ring (snapshot at $t=360$ ps) which eventually decays 
 into a pair of vortex-antivortex lines a few picoseconds later.

\begin{figure*}[!]
\centerline{\includegraphics[width=1.0\linewidth,clip]{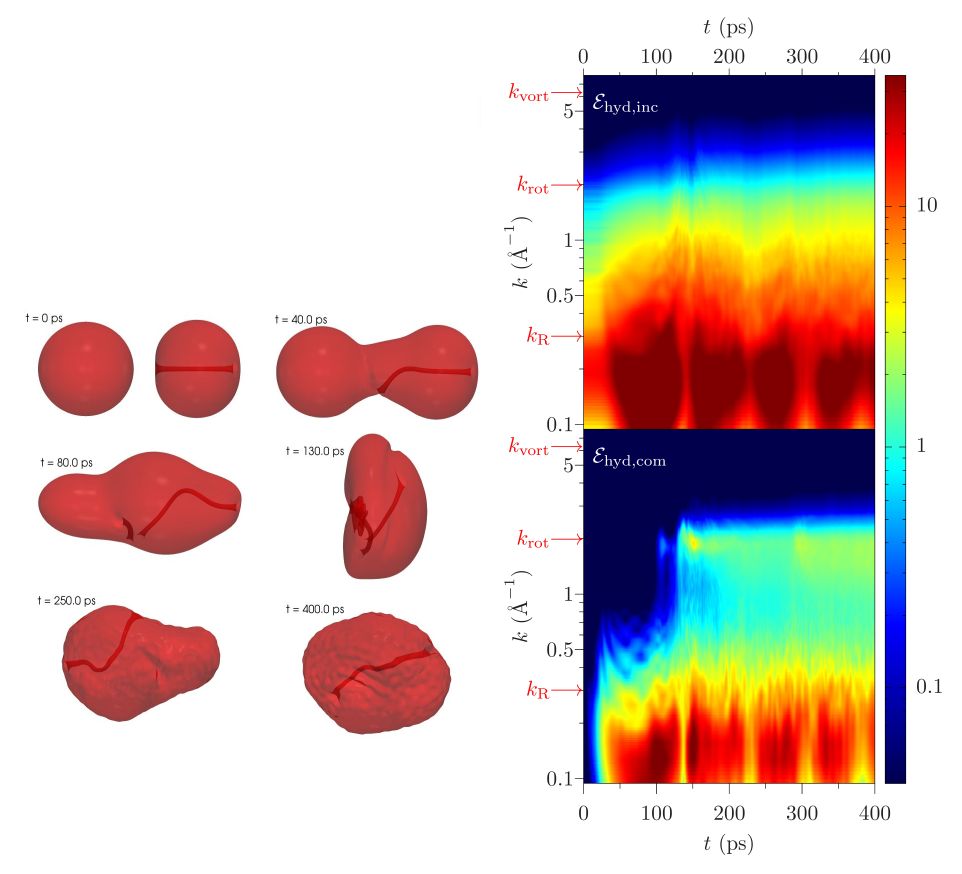}}
\caption{
Merging of two $^4$He$_{500}$ droplets, one vortex-free and the other one hosting  a vortex line along the merging direction (``corkscrew'' configuration). 
Left: Snapshots of the 3D density  at the labeled times.
Right: Energy spectrum $\mathcal{E}_{\textrm{hyd}}(k,t)$ (K\,\AA).Top, incompressible part; bottom, compressible part.
}
\label{fig5-SM}
\end{figure*}

The analysis of the energy spectrum corresponding to the anti-aligned configuration is presented in the right panel of Fig. \ref{fig4-SM}. In this case,
the roton bursts are rather weak, as they arise only from vortex-antivortex reconnections. Those at $t=488$ and 580 ps are especially marked
and produce the more intense roton peaks for this simulation. 


Finally, Fig. \ref{fig5-SM} left shows snapshots of the 3D density corresponding to the merging  of the ``corkscrew'' configuration.
Since a vortex line cannot end in the bulk of the superfluid,  it bends downwards of the merging axis in
order to hit perpendicularly  the droplet surface; it would bend upwards if the circulation of the vortex line were the opposed one. 
The energy spectrum corresponding to this  simulation is presented in Fig. \ref{fig5-SM} right.

\end{document}